\documentclass{aa}  
\pdfoutput=1
\usepackage{graphicx}
\usepackage{txfonts}
\usepackage{float}
\usepackage{booktabs}
\usepackage{placeins}
\begin{document}

   \title{TDCOSMO XV: Population analysis of lines of sight of 25 strong galaxy-galaxy lenses with extreme value statistics}
   \titlerunning{TDCOSMO XVI: Population Modeling of SL2S Lines of Sight}

   \author{Patrick R. Wells,
          \inst{1}
          \and
          Christopher D. Fassnacht\inst{1}
          \and Simon Birrer\inst{2}
          \and
          Devon Williams\inst{3}
          }

   \institute{Department of Physics and Astronomy, University of California,              Davis, CA 95616, USA
            \email{pwells@ucdavis.edu}
            \and
            Department of Physics and Astronomy, Stony Brook University, Stony Brook, NY 11794
            \and
            Department of Physics and Astronomy, University of California, Los Angeles, CA 90095, USA
        }
   \date{Received September 15, 1996; accepted March 16, 1997}


  \abstract
   {Time-delay cosmography is a technique for measuring $H_0$ with strong gravitational lensing. It requires a correction for line-of-sight perturbations, and thus it is necessary to build tools to assess populations of these lines of sight efficiently.}
   {We demonstrate the techniques necessary to analyze line-of-sight effects at a population level, and investigate whether strong lenses fall in preferably overdense environments.}
   {We analyzed a set of 25 galaxy-galaxy lens lines of sight in the Strong Lensing Legacy Survey sample using standard techniques, then performed a hierarchical analysis to constrain the population-level parameters. We introduce a new statistical model for these posteriors that may provide insight into the underlying physics of the system.}
   {We find the median value of $\kappa_{\rm{ext}}$ in the population model to be $0.033 \pm 0.010$. The median value of $\kappa_{\rm{ext}}$ for the individual lens posteriors is $0.008 \pm 0.015$. Both approaches demostrate that our systems are drawn from an overdense sample. The different results from these two approaches show the importance of population models that do not multiply the effect of our priors.}
   {}


   \maketitle
%

\section{Introduction}\label{introduction}

Time-delay cosmography is a technique for inferring the value of the Hubble constant and other cosmological parameters based on multiply imaged time-variable sources. Its independence from standard early and late Universe probes make it an essential tool in resolving the ongoing tension between those techniques. Time-delay cosmography relies on four primary ingredients. The first is time delays, which are obtained by monitoring the system over many months or years. The second is a mass model based on high-quality imaging data. The third is stellar kinematics of the lensing galaxy, which is used to break the well-known mass sheet degeneracy. The fourth, and the subject of this paper, is the external convergence (denoted $\kappa_{\rm{ext}}$), which can be thought of as the cumulative effect of all additional perturbers along the line of sight.

As with many domains of astronomy and cosmology, time-delay cosmography is increasingly big data focused. The number of known time-delay lenses has increased dramatically with surveys such as the Dark Energy Survey \citep[][hereafter DES]{des_design} and the Subaru Hyper-Suprime Cam Strategic Survey program \citep[][hereafter HSC]{hsc_design}, and is expected to increase by orders of magnitude with the Vera Rubin Observatory Legacy Survey of Space and Time \citep[][hereafter LSST]{lsst_design}.  Not only does this present unprecedented opportunities to do interesting astronomy, but it also introduces new and unique technical challenges. In particular, population modeling is an increasingly important tool for deriving constraints on interesting quantities by leveraging the statistical power of many systems. However utilizing this approach effectively is not possible without building high-quality tools that can ingest, process, and track the large amount of data required to perform the inference.

Each of the ingredients discussed above involves its own set of challenges. The $\kappa_{\rm{ext}}$ measurement is different in that the challenge is largely a problem of data management. The relative density of a given field around some lens of interest is determined by comparing the field to a large number of fields randomly drawn from some large reference survey. The data for this procedure is readily available from the various survey teams. The challenge then is building tools that are capable of doing this comparison for dozens or even hundreds of lens fields at once, while being flexible enough to allow us to evolve our techniques forward without starting from scratch. However, if this challenge is solved, this analysis serves as an excellent testing ground for building systems that do astronomy at scale.

The long term goal of the cosmography community is to provide constraints on $H_0$ with a precision comparable to that of more mature probes such as the distance ladder \citep[e.g.,][]{Riess_2022} or the cosmic microwave background \citep[e.g.,][]{planck_2020}. Being independent of these probes, cosmography is well positioned to provide an insight into the Hubble Tension \citep{Di_Valentino_2021} once this higher level of precision is realized. The combined statistical power of the population of lenses that will become available in the next decade should be sufficient to provide such a constraint, but only if we have the analysis tools to match it. In this work, we analyze a population of 25 strong galaxy-galaxy lenses from the Strong Lensing Legacy Survey (SL2S) sample. We use the number counts techniques described in \cite{0924_los} to estimate $\kappa_{\rm{ext}}$ along the line of sight to each individual lens, and then use the hierarchical techniques discussed in 
\cite{hier_los_app} to infer population parameters. We introduce a new statistical model for the resultant distributions, which provides a potential insight into the primary source of signal in $\kappa_{ext}$. This process additionally serves as a test of the ability of our techniques and software to work at scale.

In section \ref{los_basics}, we present the essentials of time-delay cosmography and the techniques used to estimate $\kappa_{\rm{ext}}$ along a given line of sight. In section \ref{los_scale}, we discuss the challenges of operationalizing this analysis to run at scale, and the tools used in the analysis presented in this work. In section \ref{data_proc}, we introduce the lens sample used in this work and discuss the choices made to estimate $\kappa_{ext}$ for the individual lines of sight. In section \ref{evs}, we introduce a novel statistical model that provides insight into the source of signal in $\kappa_{\rm{ext}}$, while in section \ref{population_techniques} we leverage these statistics to produce a population model of $\kappa_{\rm{ext}}$ for our systems. Finally, in section \ref{rad} we present and discuss the results of our hierarchical analysis, and look forward to future work on this topic. 

\section{Time-delay cosmography and line-of-sight analysis}\label{los_basics}

In this section, we present the essentials of time-delay cosmography and the associated line-of-sight analysis. For a more complete overview of time-delay cosmography and its current status, we refer the reader to \citet{tdc_overview_16}, \citet{tdc_overview_23}, and references therein, while for a more complete discussion of the line-of-sight analysis we refer the reader to \citet{0924_los} and \cite{rusu_0435}.

\subsection{$\kappa_{\rm{ext}}$ and its application to time-delay cosmography}

Time-delay cosmography relies on the strong gravitational lensing of a time-variable source to place constraints on the distance scales of the combined observer-lens-source system and, ultimately, use these constraints to derive a constraint on $H_0$. In a strongly lensed system in which multiple images are visible, the light from the various images will take different paths from the source to the observer. This difference will be directly visible when the luminosity of the source varies, as the brightness of the various images will change at different times. These techniques have been applied to strongly lensed quasars for a number of years \citep[e.g.,][]{Kundic_1997, Fassnacht_2002, Vuissoz_2008, Bonvin_2016} More recently, time-delay techniques have been applied to the supernova Refsdal \citep{refsdal_h0}, as well as cluster-scale lenses \citep{Yuting_2023}.

This time delay between two images at angular positions $\theta_A$ and $\theta_B$ of a source at (unobservable) angular position $\sigma$ can be directly related to the gravitational potential of the lens by

\begin{equation}\label{td_basic_eqn}
    \Delta t_{AB} = \frac{D_{\Delta T}}{c}[\tau(\theta_A, \sigma) - \tau(\theta_B, \sigma)],
\end{equation}

\noindent where $\tau$ is the Fermat potential of the lens given by

\begin{equation}
    \tau(\theta, \sigma) = \frac{(\overrightarrow{\theta} - \overrightarrow{\sigma})}{2} - \psi(\overrightarrow{\theta}),
\end{equation}

\noindent and $\psi(\overrightarrow{\theta})$ is the scaled lensing potential.

The cause of the time delay is the difference in the path length taken by the light of the various images plus the difference in the Shapiro delay. The quantity $D_{\Delta t}$ from Eq. \ref{td_basic_eqn} is known as the time-delay distance and is given by

\begin{equation}
    D_{\Delta t} = (1+z_d)\frac{D_{\rm{d}}D_{\rm{s}}}{D_{\rm{ds}}} \propto \frac{1}{H_0},
\end{equation}

\noindent where $D_{\rm{d}}$, $D_{\rm{s}}$, and $D_{\rm{ds}}$ are the angular diameter distances to the deflector, source, and from the deflector to the source, respectively. Given an accurate lens model and a measurement of a given time delay, it is therefore possible to measure $H_0$.

However this analysis is complicated by the fact that lenses are embedded in the Universe, and therefore surrounded by mass structures that also have an impact on the lensing observables. This effect is paramaterized with the external convergence ($\kappa_{\rm{ext}}$). The result of these perturbations is difficult to pin down, both because the mass distribution along the line of sight cannot be directly observed, and because the effect is much less significant than the effect of the primary lens. However, correcting for the effect is essential in cosmography because it propagates directly to the inferred value of $D_{\Delta t}$ by

\begin{equation}
    D_{\Delta t} = \frac{D_{\Delta t}'}{1-\kappa_{\rm{ext}}}, 
\end{equation}

\noindent where $D_{\Delta t}'$ denotes the uncorrected value of the time-delay distance. The relationship to the inferred value of the Hubble constant is similarly straightforward:

\begin{equation}
    H_0 = (1-\kappa_{\rm{ext}})H_0',
\end{equation}

\noindent where $H_0'$ represents the uncorrected value. In general, $\kappa_{\rm{ext}}$ is of order $10^{-2}$, and so failing to correct for this effect introduces bias of a few percent in the average case, up to around $10\%$ in the most extreme cases. It is worth noting that the total convergence is the combination of $\kappa_{\rm{ext}}$ and the convergence from the primary lens. In that sense, $\kappa_{\rm{ext}}$ can be thought of as the residual convergence that would be present if the primary lens was removed.

Because the underlying mass distribution in a given line of sight cannot be directly observed, we must infer it based on available data about the luminous matter in the field. The standard tools for doing this inference involve weighted number counts of galaxies within some distance of the lens. This technique has been used extensively by the TDCOSMO collaboration and its predecessors \citep[see for example][]{Fassnacht_2010, rusu_0435, 0924_los} to provide an estimate of $\kappa_{\rm{ext}}$ along a given line of sight. $\kappa_{\rm{ext}}$ can be thought of as the density of mass sheet which, if placed coplanar to the lensing galaxy, would produce the same cumulative effect as all the perturbers along the line of sight. We note that this is distinct from the internal mass sheet transformation \citep[see][]{Chen_2021, Gomer_2020}. The combination of these two effects leads to the well-known mass sheet degeneracy, which is usually broken with measurements of stellar kinematics \citep{Shajib_2023, Schneider_2013}. Generically, $\kappa_{\rm{ext}}$ results in the magnification or de-magnification of the images of the background source, but this effect is not directly measurable.

\subsection{Essentials of the technique}

In the context of lensing, $\kappa$ is a dimensionless measurement of the underlying matter distribution in units of the lensing critical density, $\Sigma_{cr}$.

\begin{equation}
    \Sigma_{cr} = \frac{c^2 D_s}{4\pi G D_s D_{ds}}
\end{equation}

In strong lensing, $\kappa > 1$ and an appropriately-placed source will be lensed into multiple images and/or an Einstein ring. For lower mass concentrations ($\kappa << 1$) lensing is instead evident by distortions in the shape of background sources, such as galaxies. Typical weak lensing techniques involve statistics based on distortions to the apparent shape of large numbers of galaxies. As a result, weak lensing analyses are typically done on much larger angular scales than is useful for the $\kappa_{\rm{ext}}$ measurement. While weak lensing analyses of $\kappa_{\rm{ext}}$ have been done \citep[e.g.,][]{wl_0435,wl_1608}, these rely on high resolution imaging (typically space-based) of the field of interest. It is unrealistic to expect such imaging for the vast majority of lens fields in future surveys, and the techniques used to do this analysis must reflect this. 

On larger scales, $\kappa$ can be thought of a measurement of the relative density of a region of space as compared to the entire Universe. An "average" field will be assigned a value of $\kappa$ near zero, while a slightly overdense field should receive a slightly positive value. For a given line of sight, we first make an empirical estimate of the density of the field. While we cannot directly observe the majority of the mass in a given field, we can use visible matter as a tracer of the underlying dark matter. In particular, a field with more galaxies is likely to have more dark matter than an identically shaped field with fewer galaxies. The relationship between luminous matter and dark matter is noisy, but we can readily estimate the amount of luminous matter in a given field with galaxy surveys.  We measured the absolute density of a given field with summary statistics computed based on the galaxies in the region. Natural summary statistics include an inverse distance summary statistic, where galaxies closer to the center of the line of sight are weighted more heavily, or a weighting based on redshift, where sources are weighted more heavily where the lensing efficiency is higher. The value of the summary statistic for a given line of sight is just the sum of the weights of the individual galaxies:

\begin{equation}
    W_{i} = \sum_{j\in gal} w_j,
\end{equation}
where $w_j$ denotes the weight for a single galaxy along the line of sight. In this context, the full posterior in $\kappa_{\rm{ext}}$ can be written as

\begin{equation}
    p(\kappa_{\rm{ext}} | \textbf{d} ) \propto p(\kappa_{\rm{ext}} | W ) p(W | \textbf{d}) \propto p(W | \textbf{d}) p(W | \kappa_{\rm{ext}}) p(\kappa_{\rm{ext}}),
\end{equation}

\noindent where the relationship between the second and third form follows from Bayes' theorem. In general, we use constraints for several summary statistics when estimating $\kappa_{\rm{ext}}$. The likelihood $p(W | \kappa_{\rm{ext}})$ cannot be written down in closed form, and we must turn to Approximate Bayesian Computation to estimate it. We seek to compare our line of sight to similar lines of sight in a simulated dataset, where values of $\kappa_{\rm{ext}}$ have already been calculated.

Usage of simulations in this context may introduce bias into the inference based on the underlying cosmology of the simulation. To control this bias, we wish to estimate the relative density of a given line of sight compared to all lines of sight in the Universe, and then find lines of sight in the simulated dataset with the same relative density to estimate $\kappa_{\rm{ext}}$.

To estimate the relative density of a given line of sight, we compute the same set of summary statistics in a large number of randomly selected fields in an appropriately large sky survey. For each random field, we compute the ratio of the summary statistic in the field of interest to the summary statistic of the given random field. The resultant distribution of ratios gives an empirical estimate of the relative density of the field compared to the Universe as a whole, so long as the reference survey is large enough to avoid sampling bias. This caveat is increasingly less of a concern. Modern surveys such as HSC and DES have hundreds to thousands of square degrees of contiguous, high quality sky coverage. In the near future LSST will image nearly all of the southern sky with many thousands of strongly lensed objects expected to be discovered. Of these thousands of systems, several hundred are expected to be suitable for time-delay cosmography \citep{lsst_lensing_forcast}. 

We then compute the same set of summary statistics in a simulated dataset with values of $\kappa_{\rm{ext}}$ computed. We normalize the resultant distribution by its median. This allows us to more directly compare these lines of sight to lines of sight of interest. More concretely, if the median of a distribution for a summary statistic in the real data is 1.2, this implies that the value of the summary statistic for our line of sight is 20\% higher than the median line of sight in our comparison survey. A line of sight from the simulated dataset with a normalized summary statistic of 1.2 is also 20\% greater than the median value (for a further discussion of the summary statistics we use, and our techniques for matching to the simulated dataset, see \citet{0924_los}).

\section{Line-of-sight analysis at scale}\label{los_scale}

A crucial aspect of the work presented here is developing and validating the tools needed to perform line-of-sight studies at scale. In \citet{0924_los}, we presented \texttt{heinlein}, a data management tool for survey datasets, and \texttt{lenskappa}, which utilized the capabilities of \texttt{heinlein} to perform a line-of-sight analysis. \texttt{lenskappa} was quite limited in that it was only capable of performing a single lens analysis at a time, and had minimal flexibility to evolve our techniques forward. In particular, \texttt{lenskappa} required weighted number count analysis for each lens to be performed individually, even if all the analyses were using the same region of the sky for comparison. This problem was magnified when performing weighted number counts in the Millennium simulation, as many more samples are required to produce a reasonable posterior.
    
The core philosophy of the $\kappa_{\rm{ext}}$ analysis is that simpler statistics can produce meaningful results when evaluated over very large datasets. In the context of this work, the correlation between the summary statistics and the quantity of interest ($\kappa_{\rm{ext}}$) is fairly weak. However, the advantage of these statistics is ease of computation. The true computational challenges of this style of analysis derive from the need to efficiently manage and query a large survey datasets. This challenge is not unique to our analysis, and building tools to efficiently solve it will be useful for a wide variety of analyses over the coming decade. We call this style of analysis "cosmological data sampling," because it requires repeatedly drawing samples from a large survey dataset. We seek to build a tool that is capable of doing this style of analysis efficiently, and allows the user to iterate and build new analyses quickly.

To this end, we introduce \texttt{cosmap}\footnote{availabile from \texttt{pip} or \url{https://github.com/PatrickRWells/cosmap}}, a Python package for defining and running "cosmological data sampling" analyses like those discussed in this work. In practice, \texttt{cosmap} can be used to apply any computation across a large survey dataset quickly and reliably. \texttt{cosmap} is an evolution of the \texttt{lenskappa} package first presented in \citet{0924_los}, and has been written from the ground up to provide an easy-to-use tool for doing analysis with big data astronomy. \texttt{cosmap} makes use of \texttt{pydantic}\footnote{\url{https://pydantic.dev}} for parameter validation and \texttt{Dask}\footnote{\url{https://www.dask.org}} to distribute work across available computing resources. Data management and result outputs can be handled by the library without user involvement, but a plugin architecture is included to modify default behavior if the user finds it insufficient for their analysis. 

Crucially, all user-defined behavior in \texttt{cosmap} is written outside of the core library. Analyses are defined as a series of transformations on the data organized as a Directed Acyclic Graph (DAG), a structure in common use in pipeline orchestration and task scheduling. Analysis parameters must be declared, and their runtime values parsed by \texttt{pydantic} to ensure correctness. These decisions ensure that failures occur early in the runtime of the program, to avoid situations where processor (and astronomer) time is wasted. 

While \texttt{lenskappa} was only capable of analyzing a single lens at a time, \texttt{cosmap} allows us to write an analysis that handles all the lenses in our sample in a single run. This saves a large amount of computation time over the previous model. The fundamentally modular nature of individual analysis definitions makes it simple to iterate on an existing analysis or define a new analysis entirely. We estimate \texttt{cosmap} saves over $95\%$ of the computational time that would be required if this analysis was done with $\texttt{lenskappa}$.

\section{Data and procedures for individual $\kappa$ measurements}\label{data_proc}
\begin{figure*}
\centering
\includegraphics[width=15cm]{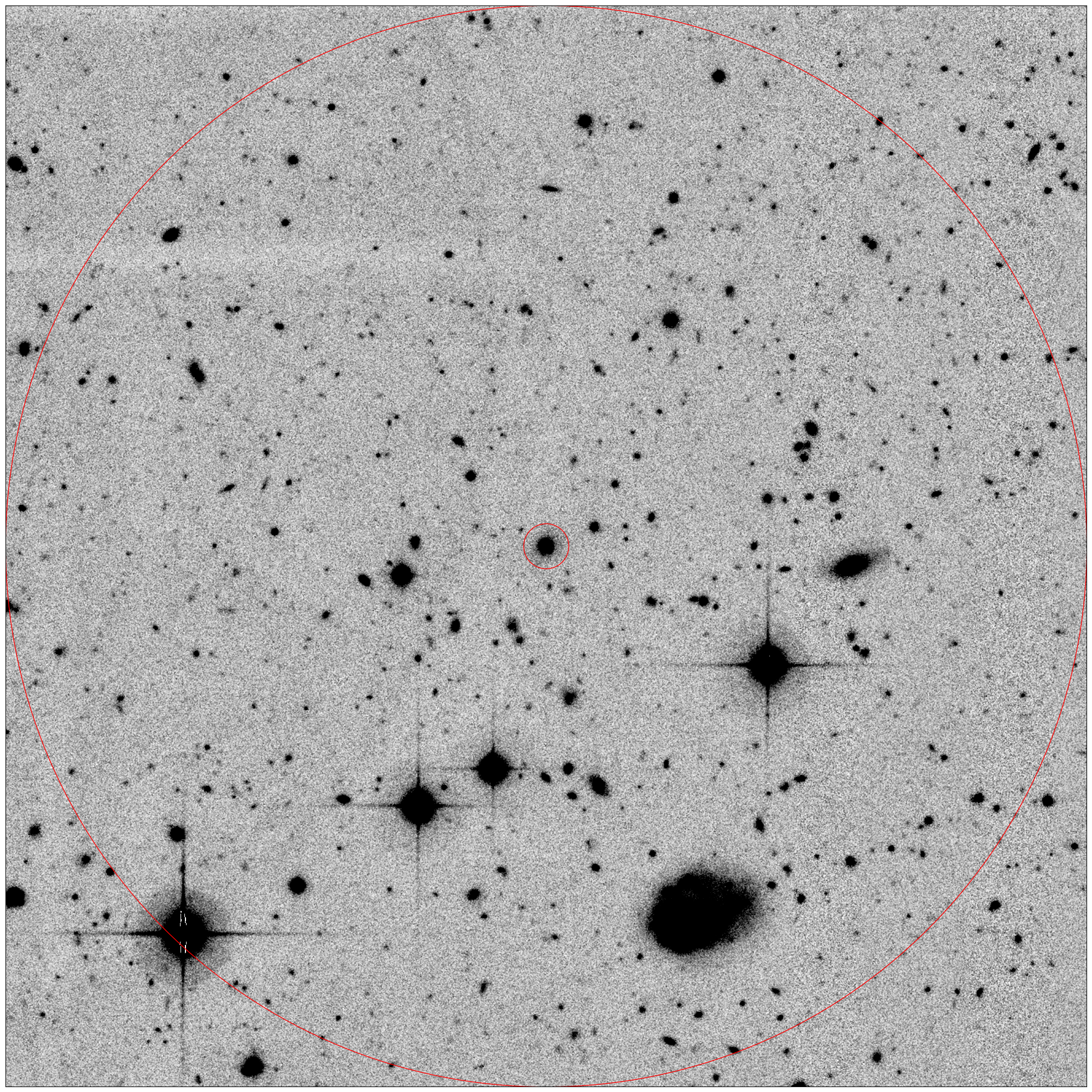}
\caption{Image showing the field around SL2SJ1405+5243. The inner and outer red circles mark the inner and outer cutoff radius we use when computing weighted number counts.}
\label{los_image} 
\end{figure*}
In this section, we discuss the dataset we use in this analysis and our procedures for performing $\kappa_{\rm{ext}}$ measurements on our individual lens lines of sight. The basic procedure used to analyze the individual lenses is identical to the procedure discussed in section \ref{los_basics}. One key difference is our ability to analyze many lenses at once, as we discussed in section \ref{los_scale}. However, we note this is a computational optimization, and does not impact the results for individual lenses.

\subsection{The CFHT Legacy Survey and the Strong Lensing Legacy Survey}

The Canada-France-Hawaii Telescope Legacy survey (hereafter CFHTLS) is a $155\: deg^2$ multiband imaging survey completed in 2012 \citep{Gwyn_2012}. After the completion of the survey, the data were re-processed with the goal of discovering strong lenses, resulting in the so-called Strong Lensing Legacy Survey \citep[][hereafter SL2S]{Cabanac_2006}. CFHTLS has been used previously to to analyze other lens lines of sight \citep[see, for example,][]{rusu_0435}. It is useful due to its depth ($i \sim 24.5$) and relatively large size (at least historically) of its wide fields.

Our sample includes 28 lenses from this survey. The sample was selected to be analogues of the kinds of systems analyzed by TDCOSMO, with the original goal of obtaining population-level constraints on the mass distribution of the lensing galaxies in the sample. The choice of lenses is discussed in more detail in TDCOSMO Collaboration (in prep). The lensing galaxies have redshifts between 0.238 and 0.884, while the source galaxies have redshifts between 1.19 and 3.39. Because our analysis relies on a high-quality galaxy catalog of the line of sight, we remove three lenses where a nearby bright star has corrupted the resultant catalog such that more than half of the field is missing galaxy photometry. Table \ref{table:lens_table} summarizes the essential information about each lens system included in this work, while Figure \ref{los_image} illustrates the typical quality of imaging data used to derive catalog products. However, we do not re-derive any catalog products, instead using the fiducial measurements performed by the survey team. 

\subsection{Individual $\kappa_{\rm{ext}}$ measurements}\label{individual_kappas}

We used the techniques discussed in Section \ref{los_basics} and in \citet{0924_los} to infer the posterior on $\kappa_{\rm{ext}}$ for each individual line of sight. At this stage, there is no information about the population-level statistics. Each line of sight is analyzed on its own, with a prior set by the Millennium simulation.

\subsubsection{Comparison field and cuts}

To compute weighted number counts for the individual lenses, we used $~50 \: deg^2$ from the CFHTLS W1 field, bound by $31^{\circ} < RA < 38.5^{\circ}$ and $-11^{\circ} < Dec < 4^{\circ}$ as a control field. We computed weighted number counts in a $120^{\prime\prime}$ aperture, and limited our counts to objects brighter than 24th magnitude in i-band. These choices were consistent with choices made in previous work \citep[see for example][]{0924_los, rusu_0435}. In particular, the magnitude limit is sufficiently bright as to be meaningfully above the survey's detection limit, while being sufficiently faint to catch all structures that are likely to contribute meaningfully to $\kappa_{\rm{ext}}$ \citep{Collett_2013}. For each lens, we also ignored objects beyond the redshift of the source quasar, and performed the same cut when comparing to the reference survey. Additionally, we removed all objects from the underlying catalog within $5"$ of the center of the field. For time-delay lenses, objects near the center of the field are typically included in the mass model explicitly, and removed during the $\kappa_{\rm{ext}}$ measurement accordingly.  

\subsubsection{Selecting Summary Statistics}

Selecting appropriate summary statistics is an important step in the analysis described here. This has been explored extensively in previous TDCOSMO and H0licow papers \citep[e.g.,][]{rusu_0435, 0924_los}. We based our result in this work on the following summary statistics:

\begin{enumerate}
    \item Pure number counts ($w_j = 1$)
    \item Inverse Distance Weighting ($w_j = 1/r_j$)
    \item Redshift-Distance Combination ($w_j = (z_s*z_j - z_j^2) / r_j)$)
\end{enumerate}

\noindent A primary challenge of this techniques is the fairly limited data that are available on individual objects in wide-field galaxy surveys. These summary statistics provide information on how much mass is clustered near the center of the line of sight, and how much mass is clustered where the lensing efficiency is high. Importantly, these summary statistics depend on quantities which are reasonably robust in modern galaxy surveys. However in general these summary statistics are poor tracers of the underlying mass distribution, as evidenced by the width of the posteriors on indiviual lines of sight. This challenge is one of the primary motivations behind combining information behind many systems into a population-level inference. More sophisticated and/or higher order summary statistics (such as two-point galaxy clustering) may provide additional useful information but are left for a future analysis.

\subsubsection{Uncertainty and comparison to the Millennium simulation}

After computing these weighted number counts, we treated the median value of the distribution as the estimate of the overdensity or underdensity. To estimate an uncertainty in this quantity, we utilized the photometric redshift uncertainties present in the underlying catalogs. We produced 1000 copies of the original line-of-sight catalog, with redshifts for each object randomly sampled from that object's photo-z PDF. We computed weighted number counts for each of the resultant catalog and treated the fractional width of the resultant distribution as the fractional uncertainty in our measurement of the median.

When computing weighted number counts in the Millennium simulation, we used the same limits described above. We used the semi-analytic galaxy catalogs of \cite{sa_catalogs}, which were shown in \cite{rusu_0435} to provide the best results for this analysis.

To match summary statistics, we selected lines of sight from the Millennium simulation that were similar in density to the lens lines of sight based on the value of the summary statistics. The values of $\kappa_{\rm{ext}}$ for each line of sight were drawn from the maps produced in \cite{hilbert_ms}, which cover the simulation in a grid with spacing between points $\approx 3.5^{\prime\prime}$. The contribution from a given line of sight was weighted by a multidimensional Gaussian centered on the distribution medians with widths set by the uncertainties discussed above. We took into account correlations between the weights when constructing this Gaussian. This is, in essence, an Approximate Bayesian Computation computation, with one key limitation. We were limited by the lines of sight available to us in the Millennium simulation, and by the computational time required to search for lines of sight matching a given lens. For the majority of lenses this is not an issue, as there were more than enough similar lines of sight in the simulation to produce posteriors that are well fit by a smooth Generalized Extreme Value (GEV) distribution. For very overdense lenses where the posterior is noisy due to a small number of matching sightlines, we widened the search Gaussian. The majority of lenses in our sample do not require this intervention, or require only a modest widening to achieve acceptable results. We note that this procedure may bias more extreme lenses towards more moderate values of $\kappa_{\rm{ext}}$, as widening the search region will naturally include more lines of sight from the central peak of the distribution.

For each lens posterior, we determinde a best-fit Generalized Extreme Value distribution with a least-squares optimizer. Using these distributions allowed us to quickly and easily draw from our priors and posteriors when sampling the population posterior. We discuss the GEV distrubtion and its interpretations in the following section. 

\section{Extreme-value statistics and applications in astronomy}\label{evs}

Extreme-value statistics describe the expected distribution of extreme values (maxima or minima) of samples drawn from a single underlying distribution. Despite relatively minimal use in astronomy, they have numerous applications in many applied disciplines. For example, extreme value statistics can be used to model the maximum daily rainfall expected over some number of consecutive days at some location. This model is crucial for engineers working to design flood-resistant infrastructure \citep[e.g.,][]{gev_hydro}.

\subsection{The generalized extreme-value distribution and subtypes}

The generalized extreme value distribution is a continuous, unimodal distribution with location parameter $\mu$, scale parameter $\sigma$, and shape parameter $\xi$ Its probability distribution is given by

\begin{equation}
\frac{1}{\sigma}t(x)^{\xi + 1} e^{-t(x)}
\end{equation}

where

\begin{equation}
t(x) =
\left\{
    \begin{array}{lr}
        \big[1+\xi\big(\frac{x-\mu}{\sigma})]^{\frac{1}{\xi}}, & \xi \neq 0\\
        \exp(-\frac{x-\mu}{\sigma}), & \xi = 0
    \end{array}
\right\}
\end{equation}

This generalized distribution is broken down into three subtypes based on the value of $\xi.$ Type I (or "Gumbel") when $\xi$ = 0, type II (or "Fréchet") when $\xi > 0$, and type III (or "Weibull") when $\xi < 0$.\footnote{Throughout this work, we report values with the standard sign convention used here. The \texttt{scipy} implementation of the GEV distribution which we use for our computational work uses the opposite convention.}

\begin{figure*}
\centering
\includegraphics[width=20cm]{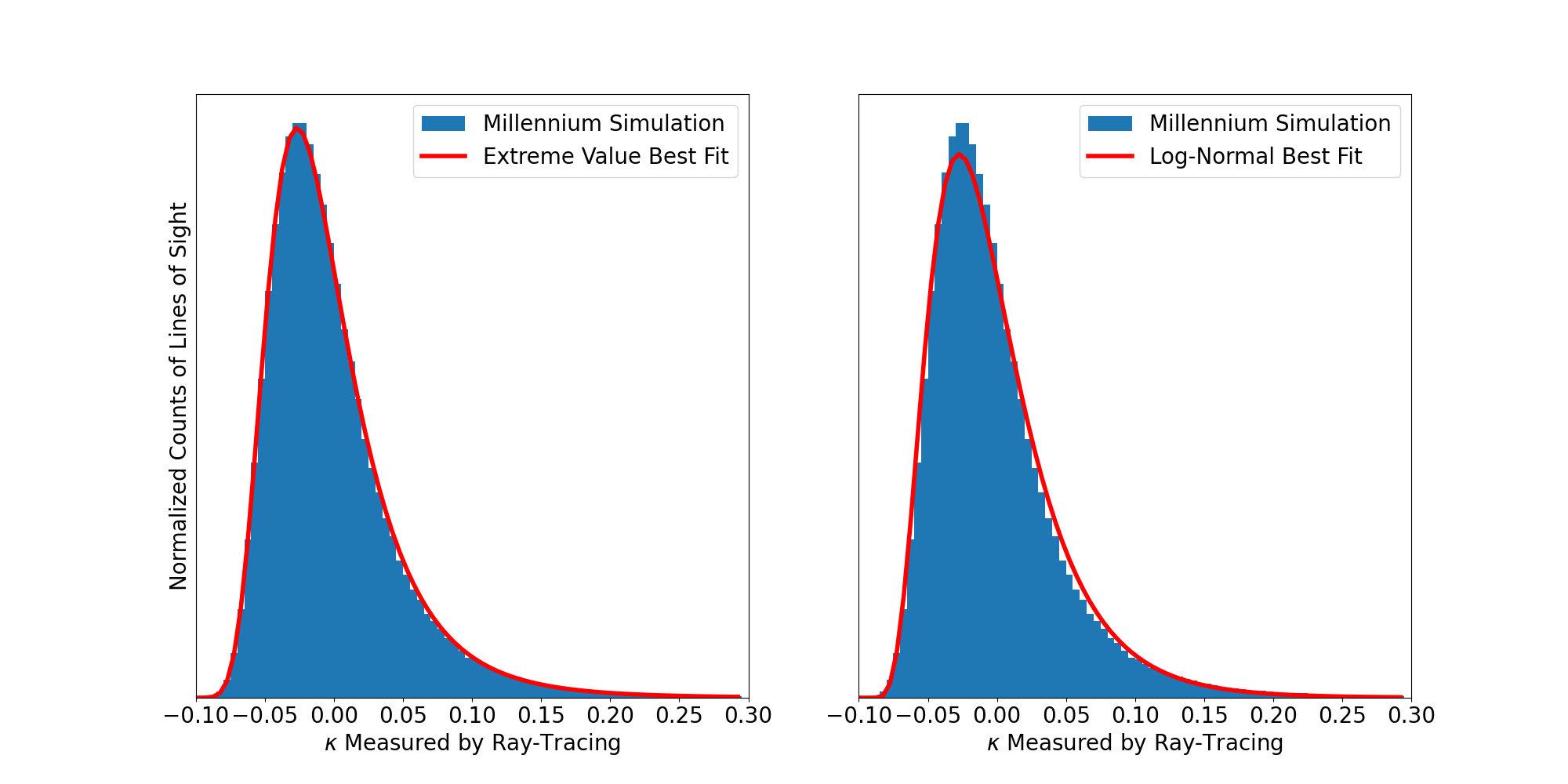}
\caption{Comparison of best-fit GEV distribution (left) and best-fit log-normal distribution(right) to the $\kappa_{\rm{ext}}$ distribution of lines of sight in the Millennium simulation at redshift z = 2.34 from \citet{hilbert_ms}. The best-fit GEV parameters are $\xi = 0.145$, $\mu = -0.0235$, and $log(\sigma) = -3.46$, while the best-fit log-normal parameters are $\mu = -0.098$ and $log(\sigma) = -0.78$}
\label{ms_gev_compare}
\end{figure*}

The Gumbel distribution typically arises when the underlying sample is normally or exponentially distributed. However for cases where the underlying distribution is bounded, the Fréchet and Weibull distributions are more appropriate. For an example of this distribution and an application to our work, see Figure \ref{ms_gev_compare}. We do not restrict our GEV fits to any of these sub-distributions at any point within this work. The corresponds to allowing the shape parameter to take on both positive and negative values, or zero as dictated by the data.

\subsection{Extreme-value statistics in astronomy}

While extreme-value statistics have found limited use in astronomy, previous work has been done showing applications in cosmic structure problems. \citet{gumbel_halo} showed analytically that the most massive halo in a given region of the Universe should follow Gumbel statistics (that is, the resultant distribution should be a GEV distribution with $\xi = 0$). \cite{gumbel_counts} demonstrated that the number of galaxies within some physical distance of a given location on the sky also followed Gumbel statistics at a wide range of scales. This second result is of particular interest, because it is nearly identical to some of the techniques discussed in section \ref{los_basics} of this work. The Gumbel distribution has additionally found use in areas such as modeling the weak lensing \citep{evs_kappa} and modeling anomalies in the Cosmic Microwave Background \citep{cmb_gumbel}. Crucially, all these analyses have implications about cosmic structure which would have a direct impact on the measured mass density in a given region of space. 

\subsection{Application to Millennium simulation and individual lens lines of sight}

The values of $\kappa$ measured in the Millennium simulation by \citet{hilbert_ms} clearly follow extreme value statistics, as can be seen in Figure \ref{ms_gev_compare}. Our posteriors on individual lenses are effectively this distribution convolved with the gaussian we use to select matching lines of sight (see Section \ref{individual_kappas}), so it is sensible they too would follow these statistics. We emphasize this choice is empirical, but the excellent fit does suggest interesting interpretations.

The distribution of $\kappa$ in the Millennium simulation and in our posteriors suggests that its value along a given line of sight may be dominated by contribution from a small number of mass structures which themselves follows extreme value statistics. The result presented in \citet{gumbel_halo} is of particular interest, because it demonstrates that the largest halo measured in a given region of the Universe should also follow extreme value statistics. This relationship suggests several interpretations that may be worth investigating in the future.

\subsubsection{$\kappa_{\rm{ext}}$ may be dominated by a single mass structure}

The appearance of extreme-value statistics in our model suggests $\kappa$ may be dominated by a single massive structure along the line of sight. This structure is not necessarily the most massive halo in the field, but may be a more moderately sized halo situated near the center of the field. In this context, the posterior on $\kappa_{\rm{ext}}$ could be interpreted as the range of mass structures (or more accurately their relative density) that are possible given some set of observables (i.e. luminous galaxies). Placing better constraints on this particular mass structure may allow us to improve the precision of the $\kappa_{\rm{ext}}$ measurement.

\subsubsection{Number counts may be able to distinguish between different halo mass models}

The appearance of extreme-value statistics both in galaxy number counts and and the underlying halo mass function suggests an interesting relationship. Number count statistics are nothing new in cosmology. Many analyses have used cluster number counts within surveys to place constraints on cosmological parameters \citep[e.g.,][]{cluster_1}. A primary limiting factor of these techniques is the measurement of the cluster masses themselves.  The power of the number counts techniques is its ease of applicability. Given some set of statistics, we can use \texttt{cosmap} to evaluate its behavior over a large region of the sky quickly and reliably. Given a set of dark matter models which make quantitatively different statements about "clustering" on small angular scales, it may be possible to quickly assess which of these models is consistent with the data available in some large galaxy survey. In this case, the constraining power of a well-measured mass structure is traded for the constraining power that derives from the scale of the dataset.

\subsection{Comparison to log-normal distribution}

A general rule of thumb in statistics is to use models with the fewest number of parameters that fit the data well. As the GEV distribution is a three parameter distribution, it is reasonable to question whether it is necessary in this context when two parameter distributions with similar shape exist. In particular, log-normal distributions have found frequent use in cosmic structure problems \citep[e.g.,][]{cpjb_lognormal, Xavier_2016} and generally provide good fits to large scale convergence and shear data \citep[e.g.,][]{Taruya_2002,Clerkin_2016}.

Our work here differs in several key ways. In particular, we are working on very small angular scales ($2'$) and only measuring out to $z\approx 2$. Nonlinear structure becomes a significant concern in this regime and it is reasonable to suggest that this may introduce complications to the standard log-normal picture.

The best-fit log-normal distribution is included in figure \ref{ms_gev_compare} in addition to the GEV fit. The log-normal fit meaningfully underestimates the peak, overestimates the decay, and slightly underestimates the tail of the emperical distribution. Both best-fit models were determined using the \texttt{stats.fit} function of \texttt{scipy} on unbinned values of $\kappa$. The best-fit log-normal distribution results in a Bayesian Information Criteria (BIC) of 219.5, while the GEV best-fit distribution yields a BIC of 174.8.

\section{Population-level environment studies in time-delay cosmography}\label{population_techniques}

Astronomy is increasingly a big data field, and time-delay cosmography is no exception. The LSST is expected to uncover many thousands of lenses in its full footprint, with many hundreds of time-delay lenses that will be suitable for cosmographic analysis \citep{lsst_lensing_forcast}. Time-delay Cosmography is still in a regime where its errors are dominated by the random variance that is to be expected from a small sample of systems. Increasingly, the community is turning to population analyses to estimate cosmological parameters as well as to more informed priors on other ingredients of the analysis such as lensing galaxy mass profiles \citep[see for example][]{hierarchical_tdcosmo}.

The measurement of $\kappa_{\rm{ext}}$ for a single lens is very much prior dominated. While easy to compute, the summary statistics we use are fairly weak tracers of the underlying matter distribution. This results in wide posteriors that often deviate only modestly from the prior except in particularly extreme cases. By combining statistical information from many lines of sight at the likelihood level, deviations of the population from the prior become more apparent. 

There is a general expectation from previous studies that lenses lie in preferably overdense lines of sight \citep[e.g.,][]{Wong_2018, Fassnacht_2010}. This is consistent with well-known work from \citet{dressler_1980} which showed that massive elliptical galaxies are more likely to be found in overdense regions. This suggests that the primary contributor to $\kappa_{\rm{ext}}$ in many lines of sight may be this group or cluster, as is suggested in \citet{Fassnacht_2010}. As will be demonstrated shortly, the analysis presented here confirms this previous work while allowing us to quantify this overdensity in a way that can be directly useful in time-delay cosmography.

\subsection {Procedures for population analysis}\label{population_procedures}

To estimate the population-level parameters of the lines of sight in our sample, we used the framework developed in \citet{hier_td_bmm} and applied in \citet{hier_los_app} to mock lines of sight. We used results from individual lines of sight to fit a distribution for the entire population. For a given value of location parameter $\mu$, scale parameter $\sigma$, and shape parameter $\xi$ the posterior takes the form:

\begin{equation}
    p(\mu, \sigma, \xi| {\textbf{d}}) \propto p(\mu, \sigma, \xi)\prod^N_{i=1} \frac{1}{M} \sum_{\kappa \in p(\kappa_i|\textbf{d})}\frac{p(\kappa | \mu, \sigma, \xi)}{p(\kappa|\Omega_{\rm{sim}})}
\end{equation}

\noindent where $p(\mu, \sigma, \xi)$ is our hyperprior on the population-level parameters, and $p(\kappa|\Omega_{sim})$ is the probability of a given value of kappa in the prior imposed by the simulation. The sum was done over 20,000 samples taken from the individual posteriors for each line of sight. We use $\sigma$ as the scale parameter in the equation above to emphasize that our target distribution is not necessarily Gaussian. We discuss our choice of target distribution in the following section.

Each value of the product term can be thought of as a likelihood for a given lens. By dividing out the prior, we avoid multiplying its effects across the population. As a result, it is reasonable to expect the population constraints to favor a more significant overdensity than a naive average of the individual posteriors. This is a statistical effect. Bayes theorem is designed as a tool for updating posteriors as new information comes in. In our case the "new information" in our analysis is that our lines of sight, looked at as a population, show significant signs of being biased when compared to the population of all lines of sight in the Universe.

\subsection{Hierarchical analysis with SL2S lenses}

Once we have posteriors for each line of sight, we move on to a hierarchical analysis of the population. We use the \texttt{emcee} Python package \citep{emcee} to sample from the posterior given in section \ref{population_procedures}.

We emphasize that the value of the prior in the Millennium simulation is not unique to a particular value of the hyperparameters, as it also depends on the redshift of the lens. It is reasonable to suggest that the redshift of sources in our sample should be included as a population parameter, but given the tight constraints on individual redshifts our sample is too small to determine this distribution.  When drawing from individual posteriors, we use the best-fit GEV distribution as provided by \texttt{scipy}.

We use a flat hyperprior with $-0.5 < \xi < 0.5$, $-0.1 < \mu < 0.3$ and $-5.0 < log(\sigma) < -2.0$. These ranges include the values of the best-fit parameters for the full population of lines of sight in the Millennium simulation.

\subsection{Interpretation of population posteriors}

When interpreting our results, it is crucial to appreciate the difference between the posterior on $\kappa_{\rm{ext}}$ produced for a single lens and the posterior on the population parameters. For a single lens, the value of $\kappa_{\rm{ext}}$ is (presumably) nearly constant across the lens system. The posterior therefore is largely a statement about our uncertainty based on the incomplete information that goes into our analysis. With better information or a more sophisticated model, it may in principle be possible to shrink the width of the posterior. However this is a posterior on only a single parameter: $\kappa_{\rm{ext}}$

However when doing a population analyses, the location, width, and shape of the population distribution are themselves parameters with associated uncertainties. These parameters are making a statement about the distribution of lines of sight in which we find strong lenses, while the issue of "incomplete information" appears in the uncertainty on the individual parameters. For the sake of intuition, it is helpful to compare the distribution produced by the parameter point estimates to the prior from the simulated dataset. However unlike the distribution for individual lines of sight, the width of this distribution has an astrophysical interpretation and may be fundamental to the population. Additionally, individual lens posteriors are prior dominated, as the data available is a fairly weak tracer of the underlying mass distribution in the field. By combining the constraining power of many lenses, it may be possible to constrain this population better than we could constrain any single lens system.

\section{Results}\label{rad}

Our results show clearly that the lines of sight in our sample are drawn from a population that is more dense than the population of all lines of sight in the Universe. 

\begin{figure*}
\centering
\includegraphics[width=18cm]{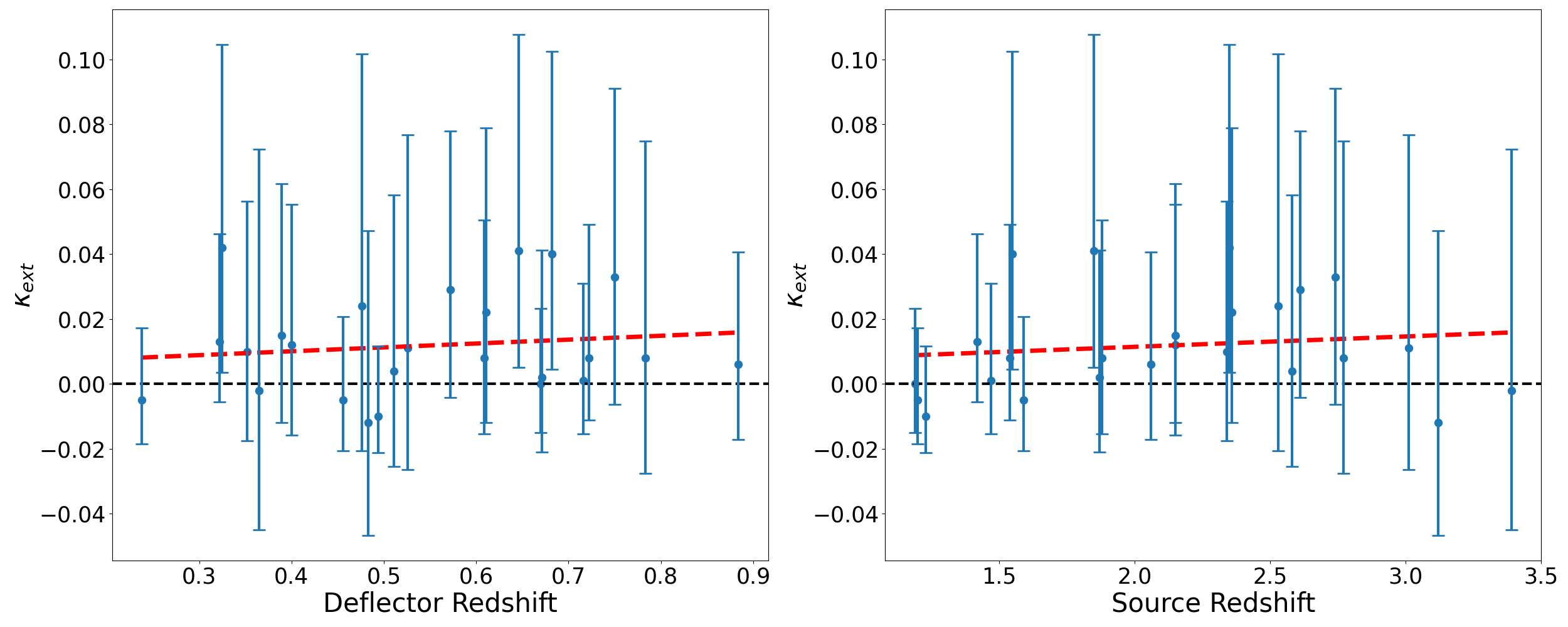}
\caption{Measured median values of $\kappa_{\rm{ext}}$ for the lenses in our sample as a function of deflector and source redshift, respectively. Error bars denote the 68\% confidence interval. The red dotted line represents the trend of the median value of $\kappa_{\rm{ext}}$}
\label{individual_results}
\end{figure*}

\subsection{Individual lines of sight}

A summary of the best-fit parameters for individual lines of sight can be found in table \ref{table:lens_table}, and a plot of our measured value of $\kappa_{\rm{ext}}$ with respect to redshift in Figure \ref{individual_results}. From the individual results, it is reasonable to suggest that our lines of sight are drawn from a population that is more dense on average than the population of all lines of sight in the Universe. Our results also demonstrate a modest trend towards greater density for longer lines of sight, but this is inconclusive.

Figure \ref{individual_results} also demonstrates the necessity of population studies. The posteriors on individual lines of sight are quite wide, and most lines of sight individually are consistent with $\kappa_{\rm{ext}} = 0$. This is a result of the fact that the summary statistics we use are fairly weak tracers of the underlying mass distribution. This amount of scatter for individual $\kappa_{\rm{ext}}$ is consistent with previous analyses on other systems.

\subsection{Population constraints}

Although individual sightlines provide limited information, our population model demonstrates clearly that our lines of sight are drawn from a biased sample. A corner plot of our $MCMC$ samples and a comparison of best-fit distributions can be found in Figure \ref{main_results}. The median value of the population distribution on $\kappa_{\rm{ext}}$ inferred by our model $\kappa_{med} = 0.033 \pm 0.010$. This result demonstrates that our lines of sight are drawn from a sample that is more dense than the general population at this redshift under the assumption the Millennium simulation provides a reasonable prior for the distribution of $\kappa$ on small scales in the Universe. However for the purposes of time-delay cosmography, the important factor is whether the median value of $\kappa_{\rm{ext}}$ is above or below zero. If our population median was exactly zero, we would anticipate that no real population-level correction from $\kappa_{\rm{ext}}$ would be necessary. Our model does not show this conclusively, but does provide evidence for this conclusion at $\approx 3 \sigma$ confidence. Our sample size is not large enough to provide strong constrxaints on the shape and scale parameters of the distribution, but this may change in the future with larger samples. In our tests, our posterior on $log(\sigma)$ remains nearly constant in the range $-8 < log(\sigma) < -5$, indicating the inability of our data to cleanly. Such a narrow posterior would indicate an extremely specific selection function, in clear contention with our knowledge of these systems. We therefore choose to cut off our hyperprior at  $log(\sigma) = -5$.

It is interesting to compare the population median to the median of the individual results (the "median of medians"). Simply averaging the individual lens posteriors results in a "population" median of $0.008 \pm 0.015$. We emphasize that this approach implicitly includes the effect of the prior once for each lens posterior, whereas the population model divides out this prior before averaging. The overdensity from our population model is not excessive, but is more significant than would be expected by a naive averaging of the results for the individual sightlines. This demonstrates that correcting for line of sight effects on a population level is necessary when performing time-delay cosmography on large samples of lenses. We remind the reader that $\kappa_{ext}$ measures the residual overdensity that remains after removing the lens and any immediate neighbors from the line of sight. The actual value of the total convergence $\kappa$ at the location of the lens itself will be quite different.

\section{Conclusions and future work}\label{conclusions}

In this paper, we have demonstrated a technique for estimating $\kappa_{\rm{ext}}$ along strong lens lines of sight at a population level and applied it to a sample of 25 lenses in the Strong Lensing Legacy Survey. This work has been built on previous work that allows us to perform $\kappa_{\rm{ext}}$ inferences on individual lenses much faster than was previously possible. We have demonstrated the infrastructure and statistical frameworks necessary to apply this technique at massive scale and provide constraints on populations of lines of sight. We have shown that the populations of lines of sight that are used in this analysis are likely drawn from a biased sample that is overdense when compared to the population of a lines of sight in the Universe, with median $\kappa_{\rm{ext}} = 0.033 \pm 0.010$ 

\subsection{Future improvements to $\kappa_{\rm{ext}}$ measurement}

The primary goal of this work is to develop and demonstrate the tools and frameworks necessary to provide constraints on large populations of lens lines of sight. While we have made much progress in this direction, there are still a number of improvements that should be made:

\subsubsection{Upgrade our simulation}

While of significant historical significance, the Millennium simulation has been surpassed in recent years by larger and more sophisticated simulations, which take the last two decades of improved understanding into account. We continue to use Millennium because of the high-resolution weak lensing maps that are available. However the \texttt{MillenniumTNG} team has produced high-resolution weak lensing maps that include the effects of baryons \citep{mtng_wl} which would be suitable for our analysis once the data products are released publicly. In particular \texttt{MillenniumTNG} has a mass resolution around one order of magnitude better than the original \texttt{Millennium} simulation, which may allow us to more cleanly map small-scale real-Universe sightlines onto equivalent simulated sightlines. 

\subsubsection{More efficient summary statistic mapping with machine learning}

Additionally, finding matching lines of sight in the simulated dataset is quite slow, as we must iterate through the entire dataset. Training a neural network to reproduce the relationship between summary statistics and $\kappa_{\rm{ext}}$ at a single redshift should be straightforward. However expanding this to encompass the entire volume of the dataset would be a much more significant challenge. Taking on this challenge may be unavoidable given the number of lenses that will discovered in LSST.

\subsubsection{Summary statistics that better target the primary mass structure}

Our work here has suggested that the primary contribution to $\kappa_{\rm{ext}}$ may be a single mass structure. Placing further constraints on this mass structure may be a way to improve the precision of the measurement. As always, a primary challenge is finding techniques that can easily be applied to a large number of systems.

\subsection{Further work on population modeling}

This work has demonstrated the techniques necessary for placing constraints on the distribution of $\kappa_{\rm{ext}}$ populations of strong lens lines of sight. While our work demonstrates with a high degree of confidence that this population of lenses fall in overdense environments, we cannot place clear constraints on the width or scale of that distribution. Additionally, we cannot say with confidence that the result derived from the SL2S sample is applicable to strong gravitational lenses as a whole. A much larger set of lenses (on the order of a few hundred) is needed to better constrain this population and boost confidence that our conclusions can be generalized. The strong lenses expected to be discovered in LSST will be an ideal sample for this type of work, and we look forward to working with these data when they become available.    

\begin{acknowledgements} 
P.R.W and C.D.F acknowledge support for this work from the National Science Foundation under Grant No. AST-1907396.

This work is partially based on observations with the NASA/ESA Hubble Space Telescope obtained at the Space Telescope Science Institute, which is operated by the Association of Universities for Research in Astronomy, Incorporated, under NASA contract NAS5-26555. Support for Program number HST-GO-17130 was provided through a grant from the STScI under NASA contract NAS5-26555.

P.W. thanks Kenneth Wong for reviewing this work prior to submission.

P.W. thanks the TDCOSMO environment working group for useful discussion throughout the preparation of this work.
\end{acknowledgements}
%
%

\bibliographystyle{aa}
\bibliography{main}
\onecolumn
\begin{appendix}

\section{Results for individual lenses in the SL2S sample}
\begin{table}[ht!]
\begin{tabular}{lrrrrrrrrrr}
\toprule
Name & RA (deg) & Dec (deg) & $z_d$ & $z_s$ & $\mu_{\kappa}$ & $\xi_{\kappa}$ & $log(\sigma_{\kappa})$ & $\kappa_{med}$ & $16\%$ & $84\%$ \\
\midrule
SL2SJ0212-0555 & 33.199 & -5.931 & 0.750 & 2.740 & 0.018 & 0.031 & -3.202 & 0.033 & -0.007 & 0.087 \\
SL2SJ0214-0405 & 33.547 & -4.084 & 0.609 & 1.880 & -0.001 & 0.179 & -3.686 & 0.008 & -0.017 & 0.036 \\
SL2SJ0217-0513 & 34.405 & -5.225 & 0.646 & 1.850 & 0.027 & 0.206 & -3.260 & 0.041 & 0.003 & 0.084 \\
SL2SJ0218-0802 & 34.505 & -8.047 & 0.884 & 2.060 & -0.003 & 0.057 & -3.738 & 0.006 & -0.017 & 0.037 \\
SL2SJ0219-0829 & 34.759 & -8.493 & 0.389 & 2.150 & 0.005 & 0.128 & -3.542 & 0.015 & -0.013 & 0.050 \\
SL2SJ0220-0949 & 35.192 & -9.824 & 0.572 & 2.610 & 0.017 & 0.002 & -3.356 & 0.029 & -0.005 & 0.077 \\
SL2SJ0225-0454 & 36.296 & -4.909 & 0.238 & 1.200 & -0.010 & 0.083 & -4.235 & -0.005 & -0.019 & 0.013 \\
SL2SJ0226-0420 & 36.544 & -4.337 & 0.494 & 1.230 & -0.014 & 0.184 & -4.383 & -0.010 & -0.022 & 0.005 \\
SL2SJ0232-0408 & 38.215 & -4.140 & 0.352 & 2.340 & -0.000 & 0.091 & -3.515 & 0.010 & -0.019 & 0.048 \\
SL2SJ0233-0438 & 38.280 & -4.644 & 0.671 & 1.870 & -0.007 & 0.153 & -3.726 & 0.002 & -0.022 & 0.030 \\
SL2SJ0848-0351 & 132.197 & -3.851 & 0.682 & 1.550 & 0.026 & 0.176 & -3.286 & 0.040 & 0.003 & 0.083 \\
SL2SJ0849-0412 & 132.290 & -4.207 & 0.722 & 1.540 & 0.001 & 0.258 & -3.825 & 0.008 & -0.014 & 0.031 \\
SL2SJ0855-0147 & 133.917 & -1.792 & 0.365 & 3.390 & -0.018 & 0.124 & -3.073 & -0.002 & -0.047 & 0.055 \\
SL2SJ0901-0259 & 135.275 & -2.985 & 0.670 & 1.190 & -0.006 & 0.102 & -4.178 & -0.000 & -0.015 & 0.019 \\
SL2SJ0904-0059 & 136.033 & -0.998 & 0.611 & 2.360 & 0.009 & 0.120 & -3.326 & 0.022 & -0.014 & 0.065 \\
SL2SJ1359+5535 & 209.957 & 55.597 & 0.783 & 2.770 & -0.006 & 0.218 & -3.270 & 0.008 & -0.030 & 0.050 \\
SL2SJ1404+5200 & 211.227 & 52.007 & 0.456 & 1.590 & -0.011 & 0.117 & -4.114 & -0.005 & -0.021 & 0.015 \\
SL2SJ1405+5243 & 211.443 & 52.720 & 0.526 & 3.010 & -0.003 & 0.139 & -3.210 & 0.011 & -0.029 & 0.059 \\
SL2SJ1406+5226 & 211.710 & 52.439 & 0.716 & 1.470 & -0.005 & 0.155 & -4.021 & 0.001 & -0.016 & 0.022 \\
SL2SJ1411+5651 & 212.904 & 56.855 & 0.322 & 1.420 & 0.006 & 0.163 & -3.917 & 0.013 & -0.007 & 0.036 \\
SL2SJ1420+5630 & 215.249 & 56.502 & 0.483 & 3.120 & -0.025 & 0.121 & -3.294 & -0.012 & -0.048 & 0.034 \\
SL2SJ1427+5516 & 216.880 & 55.279 & 0.511 & 2.580 & -0.007 & 0.172 & -3.441 & 0.004 & -0.028 & 0.041 \\
SL2SJ2203+0205 & 330.871 & 2.089 & 0.400 & 2.150 & 0.001 & 0.092 & -3.552 & 0.012 & -0.017 & 0.047 \\
SL2SJ2205+0147 & 331.279 & 1.784 & 0.476 & 2.530 & 0.007 & 0.146 & -3.045 & 0.024 & -0.023 & 0.081 \\
SL2SJ2221+0115 & 335.453 & 1.262 & 0.325 & 2.350 & 0.027 & 0.117 & -3.217 & 0.042 & 0.002 & 0.091 \\
\bottomrule
\end{tabular}
\caption{Results for individual lenses with best fit generalized extreme-value parameters and $1\sigma$ confidence for $\kappa_{\rm{ext}}$ posterior.}
\label{table:lens_table}
\end{table}
\clearpage
\section{Results of population analysis}
\begin{figure*}[ht]
\centering
\includegraphics[width=15cm]{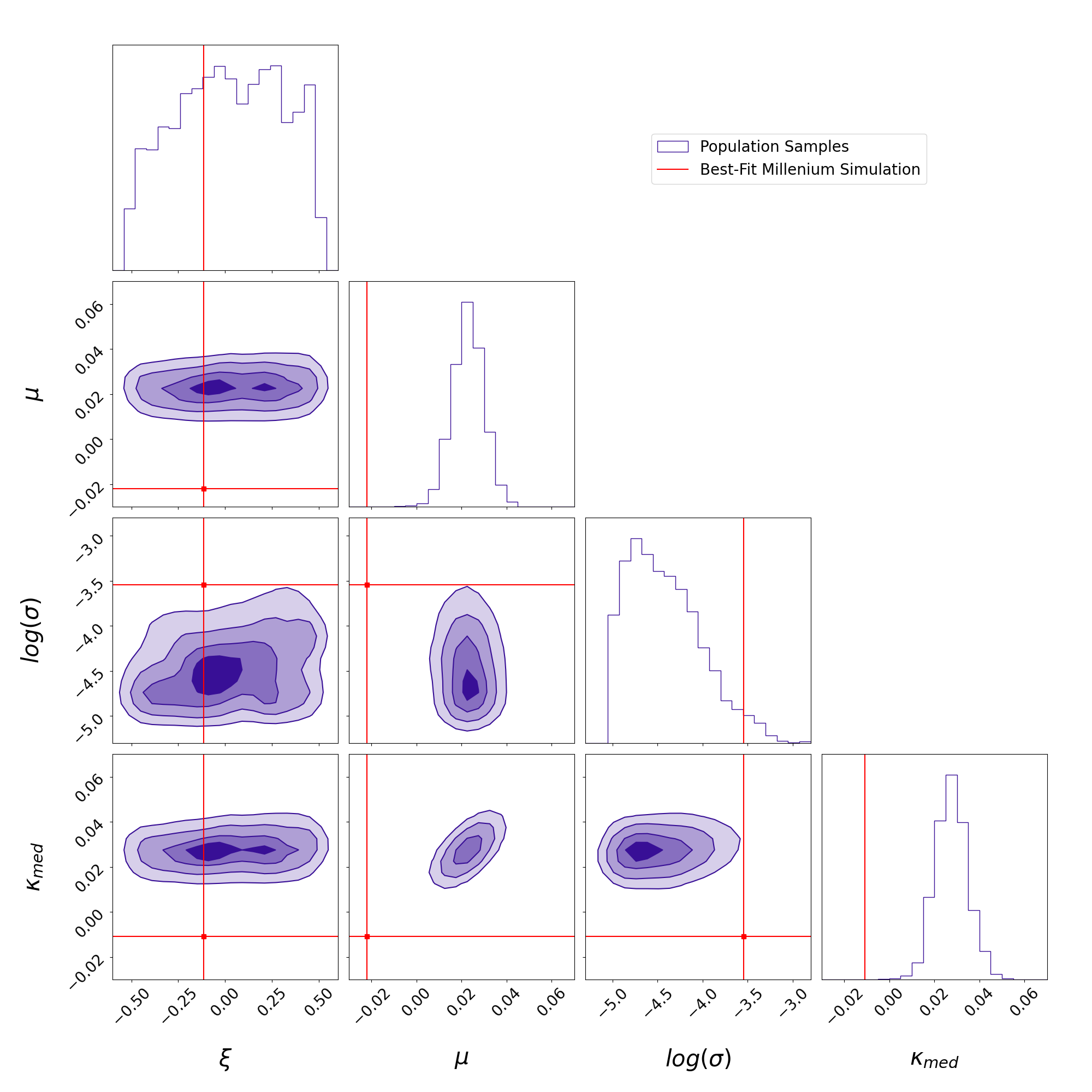}
\caption{Corner plot showing results of our MCMC, with the median value of each sample distribution included as a derived parameter. The red mark indicates the best-fit values for the entire population of lines of sight in the Millennium simulation.}
\label{main_results}
\end{figure*}

\end{appendix}
\end{document}